# Interactions between a central bubble and a surrounding bubble cluster

A-Man Zhang*[1, 2], Shi-Min Li[1], Pu Cui[1, 2], Shuai Li[1, 2], and Yun-Long Liu[1, 2]

[1]*College of Shipbuilding Engineering, Harbin Engineering University, Harbin 150001, China*
[2]*Nanhai Institute of Harbin Engineering University, Sanya, 572024, China*

**Abstract**

The interaction of multiple bubbles is a complex physical problem. A simplified case of multiple bubbles is studied theoretically with a bubble located at the center of a circular bubble cluster. All bubbles in the cluster are equally spaced and own the same initial conditions as the central bubble. The unified theory for bubble dynamics (Zhang et al. arXiv:2301.13698) is applied to model the interaction between the central bubble and the circular bubble cluster. To account for the effect of the propagation time of pressure waves, the emission source of the wave is obtained by interpolating the physical information on the time axis. An underwater explosion experiment with two bubbles of different scales is used to validate the theoretical model. The effect of the bubble cluster with a variation in scale on the pulsation characteristics of the central bubble is studied.

[1] Corresponding email: zhangaman@hrbeu.edu.cn

Bubble dynamics is a topic of great interest due to its extensive applications in underwater explosions [1-3], air-gun exploration [4-6], cavitation [7-9], and ultrasonic cleaning [10, 11]. The dynamic characteristics of a single bubble under different environments have been well developed in the past few decades [12-17]. In recent years, the dynamics of multi bubbles have been drawing increasing interest [18-22], because proper control of the characteristic parameters can promote the utilization of bubble dynamics. Experimental and numerical methods are adopted in most studies on multiple bubble dynamics. In terms of experimental research, Tomita et al. [23] quantitatively studied the coupling effect between two simultaneously generated bubbles by laser-induced bubble experiments, and found that the bubble dynamics are sensitive to the ratio of bubble scale and distance parameters. The effect of the time difference in bubble generation was systematically summarized in the works of Tomita and Sato [24], where the jet impact was found enhanced under certain conditions. Fong et al. [25] also discovered some interesting phenomena in the double-bubble interaction through spark-induced bubble experiments, such as extremely thin liquid jets. However, it is hard to accurately generate and control a bubble array in the experiments, and obtaining the pressure in the flow field requires high-precision experimental instruments of high cost. In terms of numerical research, Han et al. [26] used the boundary element method to study the coalescence of multiple bubbles. Li et al. [27] and Liu et al. [28] studied the characteristics of double-bubble interaction with the finite volume method and the finite element method. Huang et al. [29] developed a dual fast multipole boundary element method to quickly simulate the dynamic characteristics of a bubble array. Nevertheless, numerical methods have limitations in stability and accuracy due to the large deformation and splashing of bubble surfaces, especially when the number of bubbles is large. In addition, theoretical methods are also involved in some works. Bremond et al. [30] studied the mechanical properties of a hexagonal bubble cluster formed by dozens of laser bubbles, and simulated the bubble radius using the improved Rayleigh-Plesset equation. Qin et al. [31] derived analytical solutions for the pulsation characteristics of bubble clusters in different forms in incompressible fluids. In practical engineering, a large number of applications of the bubble-array exist in air-gun exploration [32-34], since an optimized arrangement for the array can fully utilize the energy of the bubbles. Zhang et al. [35] applied the particle swarm optimization method to the design of bubble-array in different environments.

The above research indicates that the dynamic characteristics of multiple bubbles are complicated, and it is hard to completely clarify the underlying mechanisms using experimental and numerical methods, especially for the pressure characteristics induced by bubbles. In terms of theoretical research, the propagation of pulsation pressure under the influence of a bubble cluster is significantly affected by the propagation time, which needs to be studied with the consideration of fluid compressibility. Thus, the research on the pressure characteristics of the flow field considering the propagation process is relative rare.

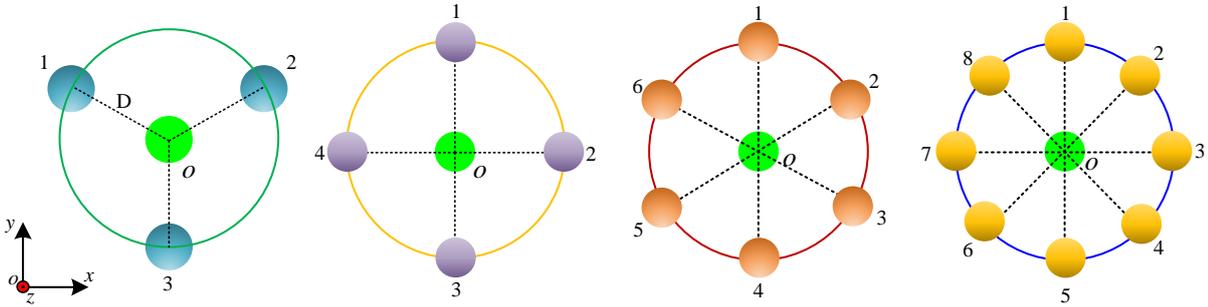

Figure 1. Arrangement form of the circular bubble cluster.

In this study, the unified theory established by Zhang et al. [36] is applied to study a simple case of multiple bubbles: a bubble is located at the center of a circular bubble cluster, and all bubbles are generated synchronously with the same initial conditions. The purpose behind this study is to, by deeply understanding the dynamics of the bubble affected by surrounding bubbles in all directions, gain insight into the pulsation characteristics of bubbles inside a bubble array and to develop new technologies in the fields of cavitation, marine exploration and underwater explosion. Figure 1 shows the form of a circular bubble cluster composed of different numbers of bubbles. The origin of the coordinates is located at the position of the central bubble (called 'bubble $o$'). The bubbles in the circular bubble cluster are equally spaced, and their distances to the central bubble at inception are 5 times the maximum bubble radius.

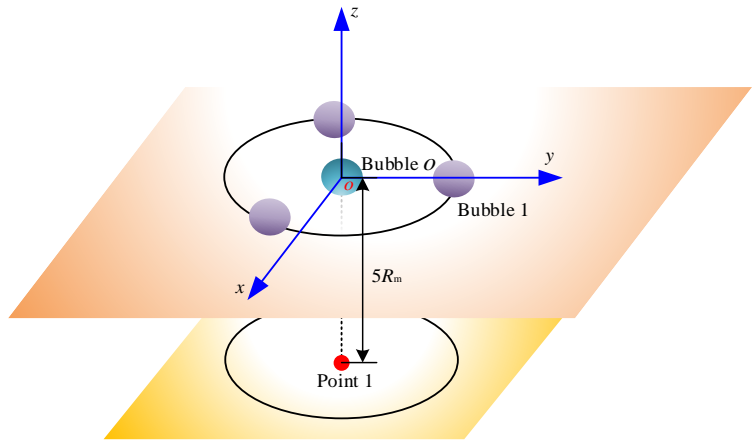

Figure 2. Schematic diagram of the location of pressure measuring points

To explore the influence of the circular bubble cluster on the pulsation characteristics of the central bubble in terms of pulsating pressure, we place the pressure measuring point in the flow field directly below bubble $o$, as shown in Figure 2. The distance between the measuring point and bubble $o$ is 5 times the maximum bubble radius. Because the bubbles in the circular bubble cluster are generated simultaneously and owns the same initial conditions, their pulsation characteristics are exactly the same. In the following study, we only show the physical characteristics of one of the bubbles which is fixed on the positive half axis of the y-axis (called 'bubble 1').

The fluids around the bubble are considered to be weakly compressible, and the gas inside the bubble obeys the adiabatic assumption. The bubble always maintains a spherical shape during the pulsation process, so according to the work of Zhang et al. [36], the unified equation of the bubble pulsation is:

$$\left(\frac{c-\dot{R}}{R}+\frac{\mathrm{d}}{\mathrm{d}t}\right)\left[\frac{R^2}{c}\left(\frac{1}{2}\dot{R}^2+\frac{1}{4}u^2+h\right)\right]=2R\dot{R}^2+R^2\ddot{R}, \quad (1)$$

where $c$ is the sound speed in water; $R$, $\dot{R}$ and $\ddot{R}$ are respectively the bubble radius, pulsation velocity and pulsation acceleration; $u$ is the value of migration velocity of the bubble; $h$ is the enthalpy [36, 37] on the bubble surface:

$$h=\frac{P_\mathrm{b}-P_\mathrm{a}}{\rho}-\frac{1}{2}\left(\frac{P_\mathrm{b}-P_\mathrm{a}}{\rho c}\right)^2+O\left(\left(\frac{P_\mathrm{b}-P_\mathrm{a}}{\rho c^2}\right)^3\right), \quad (2)$$

where $P_\mathrm{b}$ is the internal pressure of the bubble, $P_\mathrm{b}=P_0(R_0/R)^{3\kappa}$ (the subscript '0' denotes the physical quantity at the initial moment; $\kappa$ is the specific heat ratio of gas); $P_\mathrm{a}$ is the pressure of background flow field of the bubble, and its value equals to the hydrostatic pressure at infinity $P_\infty$ when the bubble is in the free field; $\rho$ is the fluid density around the bubble.

With the fourth-order Runge Kuta method, the time history of bubble radius can be obtained by solving the expanded differential form of Eq. (1):

$$\left(1+\frac{\dot{R}}{c}\right)h+\frac{R}{c}\dot{h}+\frac{1}{4}\left(1+\frac{\dot{R}}{c}\right)u^2+\frac{R}{2c}u\dot{u}=\frac{3}{2}\left(1-\frac{\dot{R}}{3c}\right)\dot{R}^2+\left(1-\frac{\dot{R}}{c}\right)R\ddot{R}, \quad (3)$$

Compared with the Keller-Mixis equation [38] the effect of bubble migration on the pulsation characteristics is considered in Eq. (3); thus, the bubble migration needs to be solved separately. Base on the conservation of momentum for the bubble [36], the migration of the bubble is obtained by the following equation:

$$C_\mathrm{a}R\dot{\boldsymbol{u}}+3C_\mathrm{a}\dot{R}\boldsymbol{u}+\frac{R}{\rho}\nabla P_\mathrm{a}+\frac{3}{8}C_\mathrm{d}\mathbb{S}(\boldsymbol{u})=0, \quad (4)$$

where $C_\mathrm{a}$ and $C_\mathrm{d}$ are the additional mass coefficient and drag coefficient of the bubble, respectively; $\nabla P_\mathrm{a}$ represents the pressure gradient of background flow field of the bubble, and it is the hydrostatic pressure gradient in the free field; $\mathbb{S}()$ is a special defined symbol, $\mathbb{S}(\cdot)=(\cdot)|\cdot|$.

Combine Eqs. (1) and (4), the time history of bubble radius and displacement can be obtained. The pressure of flow field [36] induced by bubble pulsation can be calculated by the following equation:

$$P(\bm{r},t) = P_{\text{a}} + \rho\left\{\frac{R}{|\bm{r}|}\left(h+\frac{1}{2}\dot{R}^2\right)-\frac{1}{2}\frac{R^2}{|\bm{r}|^4}\left[R\dot{R}+\frac{|\bm{r}|-R}{C}\left(h+\frac{1}{2}\dot{R}^2\right)\right]^2\right\}\Bigg|_{\left(R,t-\frac{|\bm{r}|-R}{C}\right)}, \quad (5)$$

where $\bm{r}$ is the position vector of the measuring point in the flow field. Note that we interpolate the physical information of the bubble on the time axis to take into account the propagation process of the pulsating pressure, that is, the physical information in the flow field at $t$ is induced by the bubble at $t-(|\bm{r}|-R)/c$.

To model the interaction of multiple bubbles, the background flow field of the bubble needs to be corrected, that is, the pressure field induced by other bubbles are superimposed on $P_{\text{a}}$. The pressure of the background flow field of the bubble [36] induced by bubble $N$ are given as below:

$$P_{\text{a}}(\bm{o},t) = P_{\infty} - \rho\sum_{N=1,L}\left(-\frac{R_N}{|\bm{o}_N-\bm{o}|}\left(h+\frac{1}{2}\dot{R}_N^2\right)\Bigg|_{(R_N,t_N)}\right)$$
$$-\frac{1}{2}\rho\left|\sum_{N=1,L}\left(-\frac{\bm{o}_N-\bm{o}}{|\bm{o}_N-\bm{o}|^3}\left[\frac{R_N}{C}(|\bm{o}_N-\bm{o}|-R_N)\left(h+\frac{1}{2}\dot{R}_N^2\right)+R_N^2\dot{R}_N\right]\Bigg|_{(R_N,t_N)}\right)\right|^2, \quad (6)$$

where $\bm{o}$ is the position vector of the studied bubble; $\bm{o}_N$ is the position vector of bubble $N$; $L$ equals to the total number of bubbles minus one; $t_N$ is the time when bubble $N$ causes disturbance to measuring point in the flow field, $t_N = t-(|\bm{o}_N-\bm{o}|-R_N)/c$.

To make the research universal, the calculation results would be dimensionless in the subsequent discussion unless specific stated. The maximum radius of the bubble $R_{\text{m}}$, the fluid density around the bubble $\rho$ and the hydrostatic pressure at infinity $P_{\infty}$ are regarded as reference physical quantities. The scale of the bubble can be characterized by the buoyancy parameter $\delta$ ( $\delta = \sqrt{\rho g R_{\text{m}}/P_{\infty}}$ ). Firstly, we illustrate the effect of fluid compressibility with a simple case of double-bubble, and compare the evolution of the bubble radius in the present theory with the results without fluid compressibility and computed by the boundary integration method (BIM) [39-41], as shown in Figure 3. The incompressible theory is achieved by letting $C$ tend to infinity in Eqs. (1) - (6), and the BIM simulation is conducted by solving the boundary integration Eq. (7) with the Bernoulli Eq. (8) as the dynamic boundary condition:

$$\lambda\varphi = \iint_s [G\frac{\partial\varphi}{\partial n}-\varphi\frac{\partial G}{\partial n}]\text{d}s, \quad (7)$$

$$\frac{\text{d}\varphi}{\text{d}t} = 1-\frac{1}{2}|\nabla\varphi|^2+\bm{v}_{\text{DPM}}\cdot\nabla\varphi-\delta^2 z-\varepsilon\left(\frac{V_0}{V}\right)^{\kappa}, \quad (8)$$

where $\varphi$ is velocity potential; $\lambda$ is the solid angle; $G$ is Green function; $s$ is the bubble surface;

$v_{DPM}$ denotes the velocity of nodes on the bubble surface obtained by the density optimization method [42]; $\varepsilon$ is the dimensionless inner bubble pressure at inception; $V_0$ and $V$ are the bubble volume at the initial and current moment, respectively.

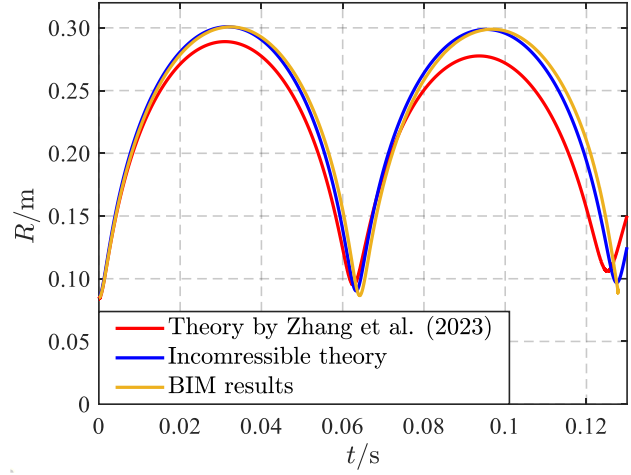

Figure 3. Comparison of the bubble radius in the present theory with the incompressible theory and boundary integral method ($C_a = C_d = 0.5$).

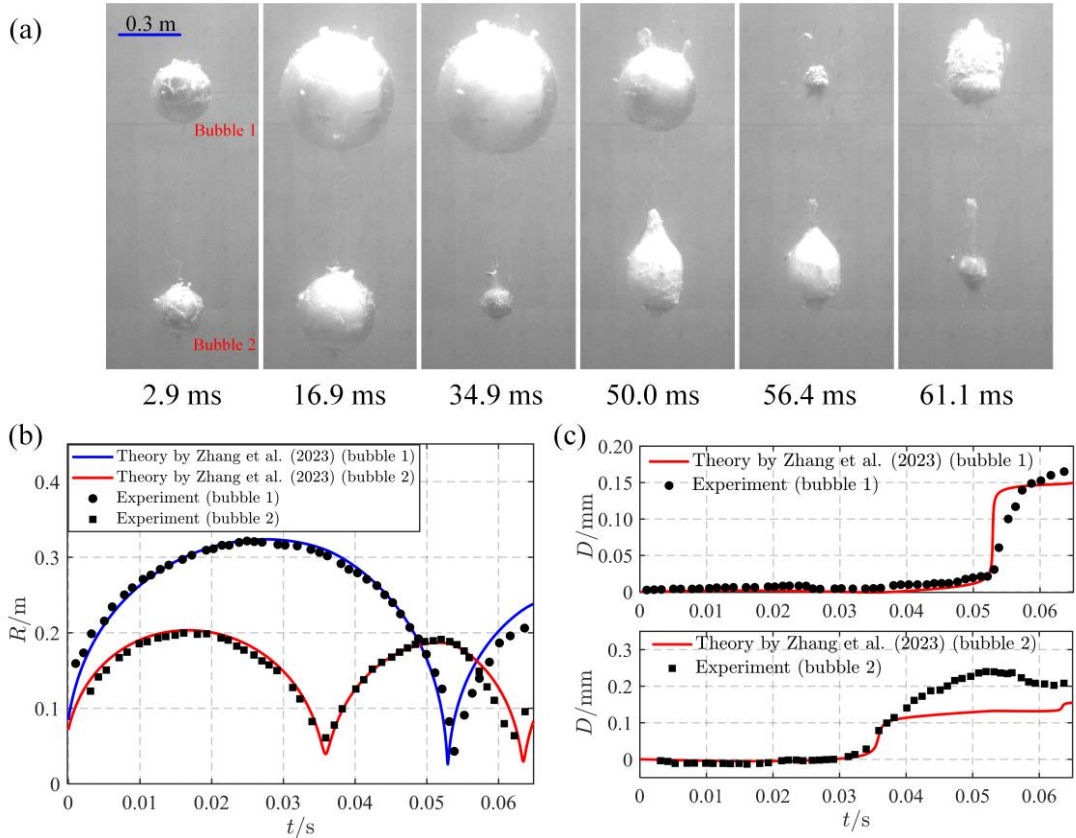

Figure 4. Comparison of the bubble dynamics between theoretical and experimental results for two underwater explosion bubbles (**a**) experimental images at typical moments (**b**) bubble radius (**c**) bubble displacement.

In Figure 3, two identical bubbles are set horizontally and generated at the same time, with the dimensionless distance of 6 and the buoyancy parameter $\delta$ of 0.15. The dimensionless initial conditions for bubble pulsation are: $R_0=0.28$, $\dot{R}_0=0$, $P_0=18$. The most obvious influence of fluid compressibility on the temporal evolution of the bubble radius is the energy loss in the bubbles. The maximum bubble radius in the first cycle in the incompressible theory and BIM is significantly larger than the present theory, and the bubble expands to the same size in the second cycle. In addition, the bubble modeled by BIM contracts more slowly at the end of collapse than in the incompressible theory although the discrepancy is not obvious. This is the result of the asymmetric deformation of bubbles delaying the velocity of bubble collapse. In general, the fluid compressibility is important for predicting the pulsation characteristics of bubbles, especially near the minimum bubble volume.

To further validate the multi-bubble coupling effect in the present theory, we compare the theoretical results with an underwater explosion experiment, as shown in Figure 4. The experimental results are displayed in the dimensional form. The underwater explosion is conducted in a 4×4×4 m³ cubic water tank and two hexogen explosives suspended vertically are detonated simultaneously with the equivalent of 8.5 g and 2.5 g. The upper bubble with the larger volume is defined as bubble 1 and the other one as bubble 2, as shown in Figure 4a. The initial depth of bubble 1 is 3.6 m, and bubble 2 is directly below bubble 1 at inception with the vertical distance of 1.2 m. To obtain the initial conditions of the bubble pulsation in the theoretical calculation, we integrate the pulsation Eq. (1) forward according to the maximum radius of the bubble in the experiment. Detailed implementation details can be found in the works of Wang [13]. Here, we directly give the initial conditions of Bubble 1 and Bubble 2: $R_{01}=0.084$ m, $R_{02}=0.072$ m, $\dot{R}_{01}=60$ m/s, $\dot{R}_{02}=40$ m/s, $P_{01}=P_{02}=0.48$ MPa, $C_a=1.0$, $C_d=0.5$. The theoretical results well reproduce the time history of the bubble radius in the experiment, as shown in Figure 4b. The computed vertical displacement of the two bubbles is compared with the experimental values, as illustrated in Figure 4c. The migration of the bubbles in theory also agrees well with experiments in general, except that the non-spherical collapse of the bubbles reduces the migration speed near the moment of the minimum bubble volume for bubble 1 and the jetting behavior leads to the oscillation of displacement and larger amplitude of migration for bubble 2. The assumption of the spherical bubble results in the difference of the bubble migration between the present theory and experiment, but overall, the computed trend of bubble migration is consistent with experimental results and the amplitude differences are acceptable.

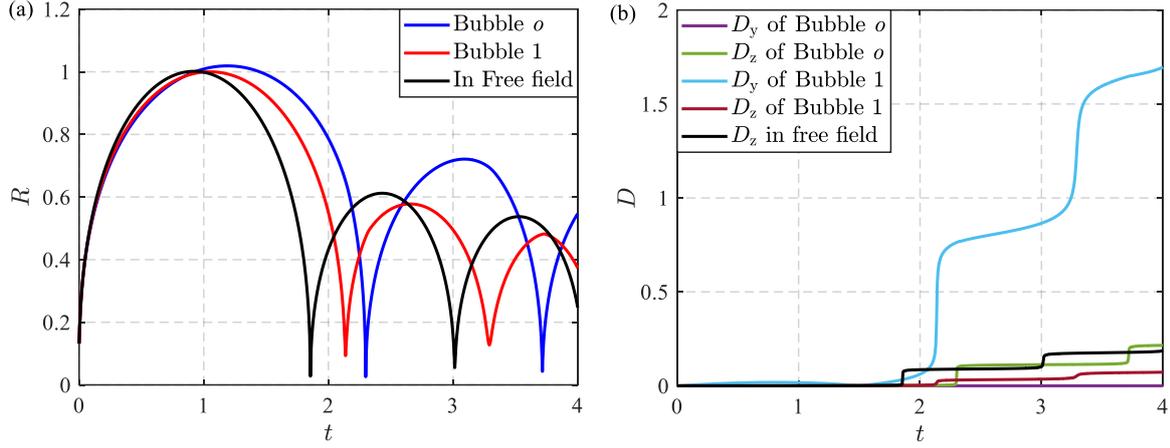

Figure 5. The radius and displacement of bubble *o* and bubble 1 under the influence of a circular cluster composed of three bubbles (**a**) bubble radius (**b**) bubble displacement

In Figure 5, the temporal evolution of the bubble radius and migration is provided for bubble *o* and bubble 1 when bubble *o* is surrounded by a cluster of three bubbles, and that of bubble *o* in the free field is also plotted. To reduce the effect of non-spherical deformation of bubbles induced by buoyancy effect, the buoyancy parameter $\delta$ is set to be small with the value of 0.04. The dimensionless initial conditions of the bubble pulsation are: $R_0=0.13$, $\dot{R}_0=18$, $P_0=12$, with the initial Mach number of the bubble-wall ($\dot{R}_0/c$) is 0.12. The oscillation of bubble *o* is hindered by the simultaneous oscillation of surrounding bubbles in the cluster, so its pulsation period is largest, as shown in Figure 5a. The pulsation of bubble 1 is only hindered by other bubbles on one side, and the fluid flow on the other side is relatively free, causing the pulsation period smaller than bubble *o* but larger than in the free field. Since the collapse velocity of bubble *o* is smaller than that of bubble 1, the energy loss of bubble *o* caused by liquid compressibility at the end of first cycle is smaller, resulting in a larger bubble volume in the second cycle. Note that the contract of bubble *o* is accelerated in the late stage of collapse compared to the expansion stage. This is due to the fact that the bubbles in the cluster begin to expand before bubble *o* reaches its minimum volume, causing the fluids around bubble *o* to accelerate to flow toward bubble *o*. Figure 4b shows the time histories of the migration magnitude of bubbles *D* in the y-direction and z-direction. The flow field around bubble *o* in the *x-y* plane remains symmetrical throughout the pulsation process, so there is no horizontal migration of bubble *o*. The migration of bubble 1 towards bubble *o* is most pronounced because bubble 1 is attracted by bubble *o* and the other two bubbles in the cluster. Compared to $D_y$ of bubble 1, the *z*-direction migration of all bubbles can be negligible due to the small buoyancy parameter.

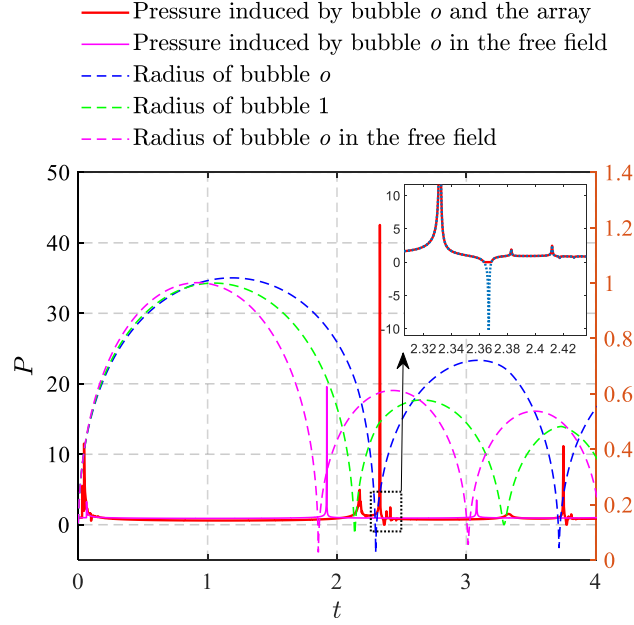

Figure 6. The pressure evolution of the measuring point under the influence of the central bubble and a circular cluster composed of three bubbles ($C_a$=1.0, $C_d$=0.5).

Figure 6 illustrates the pressure time-history curve of the measurement point in the flow field for the case of three-bubble circular cluster, and the bubble radius is also plotted in the form of dotted lines to clarify the source of the pressure peak. Likewise, to analyze the effect of the circular bubble cluster, the pressure curve induced by bubble $o$ in the free field is also given, as shown by the pink solid line. As the bubble contracts to its smallest volume at the end of each cycle, it radiates energy outward, inducing the noticeable pressure peak. Compared with the situation without the influence of the circular bubble cluster, the contraction of bubble 1 is extenuated by other contracting bubbles, causing the smaller pressure peak caused by bubble 1. Nevertheless, the opposite is true for bubble $o$. Since the collapse of bubble $o$ at the end of the first period is accelerated by the surrounding expanding bubbles, the pressure of the flow field induced by bubble $o$ becomes very large, as shown by the pressure peak near the minimum volume of bubble $o$, much larger than that induced by bubble $o$ alone. Subsequently, several small peaks appear on the pressure curve, as shown in the zoomed-in view of the black dotted box, which is the result of the reflection and superimposition of the pressure wave. The time interpolation method in the present theory reproduces this feature. Note that the reflection of pressure waves induced by the bubble $o$ on the surfaces of surrounding bubbles induces a negative pressure, similar to the situation of the bubble near a free surface [43]. The negative pressure is even lower than the saturated vapor pressure, as illustrated by the blue dotted line. We accounted for this negative pressure using the truncated model [44], letting the saturated vapor pressure be the lower limit of the pressure value. The characteristics of the pressure wave at the end of the second period are similar to that at the end of the first period, and would not be repeated here.

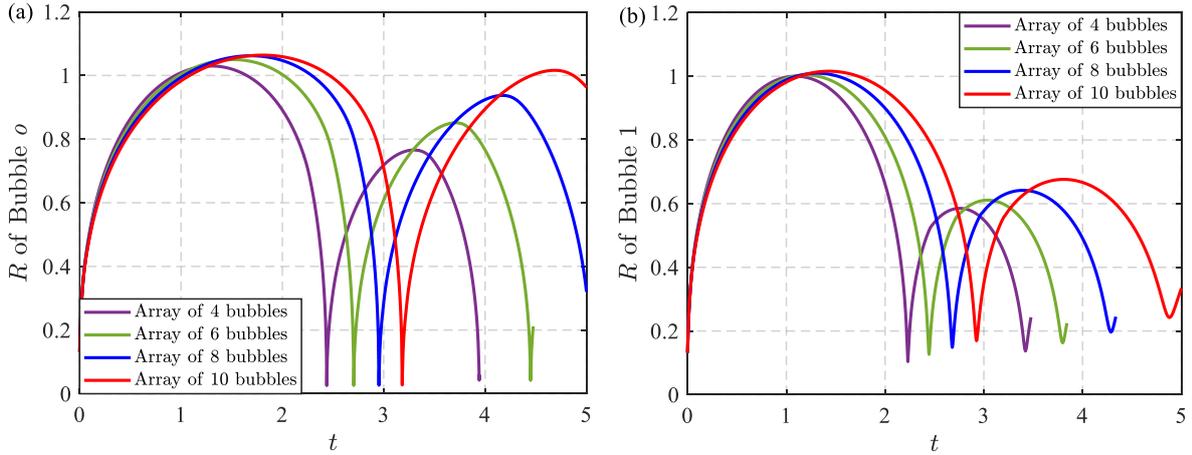

Figure 7. Time history curves of the radius for (**a**) bubble *o* and (**b**) bubble 1 under different scales of the bubble cluster.

Furtherly, we discuss the pulsation characteristics of bubble *o* under the influence of larger-scale clusters. Figure 7 shows the time history curves of radius for bubble *o* (Figure 7a) and bubble 1 (Figure 7b) under the influence of circular clusters consisting of 4, 6, 8, and 10 bubbles. As the scale of the bubble cluster increases, the period of the bubbles gradually becomes larger because the mutual hindrance among bubbles becomes stronger. Compared with Figure 5, the accelerated contraction of bubble *o* at the end of the collapse becomes more obvious, since more bubbles are in expanding before bubble *o* collapses to the minimum volume. As the number of bubbles increases, the energy loss of bubble *o* at the end of the first cycle gradually decreases, with the larger bubble volume in the second cycle. Similar features about the bubble volume and period also occur for bubble 1, as shown in Figure 7b.

Figure 8 compares the pressure of the measuring point in the flow field under different spatial scales of the circular bubble cluster. When the spatial scale of the bubble cluster becomes larger, more pressure peaks appear after the pressure wave induced by bubble *o*. This is due to the fact that the distances from one bubble to others are various, causing the pulsating pressure waves to superimpose with each other and reflect on different bubble surfaces. The magnitude and number of these pressure peaks increase with the increasing number of bubbles. In addition, it is worth mentioning that the pressure peak induced by bubble *o*, as shown by the maximum value of each curve in Figure 8, does not increase with the increasing spatial scale of the bubble cluster. More bubbles in the circular cluster lead to stronger ghost reflections, which weaken the pressure wave induced by bubble *o* to a greater extent. It is worth mentioning that the pulsation pressure peak for three-bubble circular cluster ($P = 42.5$) is smaller than four-bubble cluster ($P = 43.2$). Thus, it is concluded that the peak of pulsation pressure is highest as the circular cluster is composed by four bubbles among three to ten bubbles.

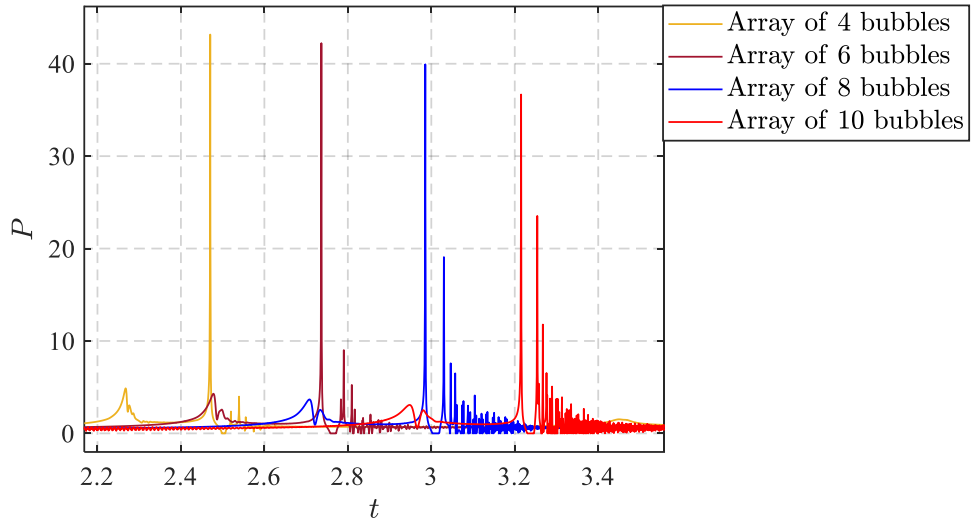

Figure 8. Pressure curves of the measuring point under different scales of the bubble cluster.

In this paper, the unified theory established in the work of Zhang et al. [36] is applied to model the dynamics of a circular bubble cluster and a bubble placed at the center. All bubbles are generated synchronously and own the same initial conditions. An experiment of two underwater explosion bubbles was compared with the theoretical results and great agreement was obtained. Under the influence of the circular bubble cluster, the period of the central bubble is obviously prolonged and the contraction is accelerated at the end of the collapse. At the measuring point directly below the central bubble, a high pulsating pressure peak is observed, which is much larger than that induced by a bubble in a free field. As the spatial scale of the bubble cluster increases, the size and pulsation period of the bubbles become larger, but the pulsation pressure induced by the central bubble is weakened due to the stronger ghost reflection, resulting in the decrease in the pressure peak. Among the circular clusters consisting of three to ten bubbles, the pulsation pressure peak of the central bubble is the highest for the cluster with four bubbles.

**Acknowledgements**

This work is funded by the National Natural Science Foundation of China (51925904, 52088102), the National Key R&D Program of China (2022YFC2803500, 2018YFC0308900), Finance Science and Technology Project of Hainan Province (ZDKJ2021020), the Heilongjiang Provincial Natural Science Foundation of China (YQ2022E017) and the Xplore Prize.